\documentclass[manuscript]{aastex}
 
\usepackage{graphicx}
\usepackage{amsmath} 
\usepackage{pstcol}

\shorttitle{Momentum transport from current-driven reconnection }
\shortauthors{Ebrahimi et al.}

\begin{document}

\title{Momentum transport from current-driven reconnection in astrophysical disks}

\author{F. Ebrahimi$^1$ and S. C. Prager$^2$}
\affil{$^1$ Space Science Center, University of New Hampshire \\$^2$ Princeton Plasma Physics Laboratory \\ 
and Center for Magnetic-Self Organization in Laboratory and Astrophysical Plasmas}

\begin{abstract}    
Current-driven reconnection is investigated as a possible mechanism 
for angular momentum transport in astrophysical disks.  
A theoretical and computational study of angular momentum 
transport from current-driven magnetohydrodynamic instabilities 
is performed.  It is found that both a single resistive tearing 
instability and an ideal instability can transport momentum in
 the presence of azimuthal Keplerian flow.  The structure of 
the Maxwell stress is examined for a single mode through 
analytic quasilinear theory and computation.  Full nonlinear 
multiple mode computation shows that a global Maxwell stress 
causes significant momentum transport.

\end{abstract}
\section{{\Large\textbf{Introduction}}} 

As matter accretes from astrophysical disks onto a central body 
(such as protostars, neutron stars, and black holes)
 angular momentum is rapidly transported outward.  
This redistribution of angular momentum is far too rapid 
to be explained by collisional viscosity.  It is believed      
that turbulence initiated and sustained by 
magnetohydrodynamic (MHD) plasma instabilities can be 
responsible for the enhancement of effective viscosity 
in accretion disks (Shakura \& Sunyaev 1973; Pringle 1981).~\nocite{shakura73,Pringle81} 
Disks appear to be hydrodynamically 
linearly stable according to the Rayleigh criterion.  
However, it is not clear whether through a nonlinear 
process a hydrodynamically stable flow can become 
turbulent, a topic that has been debated vigorously.  
Recently, it has been shown experimentally that 
hydrodynamic turbulence in astrophysically relevant 
flows at Reynolds numbers up to several million cannot        
provide the required effective viscosity 
(Ji, Burin, Schartman \& Goodman 2006).~\nocite{ji06} 

The ineffectiveness of hydrodynamic turbulence in 
transporting momentum has motivated the search 
for other instabilities that could yield enhanced 
viscosity.  A candidate for such instability is 
the magnetorotational instability (MRI).  The MRI, 
originally studied by Velikhov 1959 and 
Chandrasekhar 1960,~\nocite{velikhov59,chandrasekkhar61}  
was analyzed for astrophysical disks by Balbus    
and Hawley in the early 1990s (Balbus \& Hawley 1991).~\nocite{balbus91} 
They showed that  
a differentially rotating disk in the presence of
 weak magnetic field is MHD unstable.  Thus, 
under appropriate conditions, the MRI can lead 
to turbulent momentum transport in disks (see, e. g., 
Balbus \& Hawley 1998, and the references
therein).~\nocite{balbus98_rmp}    
The momentum transport arises from Maxwell or 
Reynolds fluid stresses.  In this paper, 
we investigate whether those stresses and 
momentum transport can be provided by a 
current-driven MHD instability, rather than 
through the flow-driven MRI.  In particular, 
we consider in depth a tearing instability, 
which is a resistive MHD instability, that 
causes magnetic reconnection and can persist 
in a strong magnetic field (in regions in which the 
MRI could be stabilized by the strong field).

Magnetic fields have long been considered to be prevalent in some 
disk regions, as well as to play a key role in jets emanating from 
disk regions.  Observational evidence is in hand for the existence 
of strong magnetic fields in disks, although the observations are 
limited. In recent years, direct measurements have shown strong 
evidence of magnetic field in protostellar disks.  Existence of 
a significant azimuthal magnetic field of about 1 kG has been 
confirmed by observations around FU Orionis in the innermost 
regions of its accretion disk (Donati et al. 2005). The complex 
topology of magnetic field around stars during their formation 
has also be confirmed through observation and extrapolation of 
the reconstructed surface magnetic field.  Recent  observation 
and analysis of magnetic field on the surface of classical 
T Tauri stars (cTTS) suggest strong octopolar field 
($\approx$ 1.2 kG) and smaller dipolar field 
($\approx$ 0.35 kG) (Donati et al. 2007).   

There is some indirect evidence of strong magnetic field 
in protosolar nebula based on the magnetization of primitive meteorites.
It was suggested that an MHD nebular dynamo can 
generate a magnetic field with intensity 
of up to 10G at the distance of a few astronomical units from the Sun 
(Levy 1978)\nocite{levy78}, and was shown that this 
magnetic field can produce flares 
in a nebular corona (Levy \& Araki 1989)\nocite{levya}, which may account for 
the magnetized meteorites. 

The source of large-scale magnetic      
fields of cTTS could be the combination of fossil field 
(from interstellar media) and field generated through 
dynamo action (Donati et al. 2007). However, the source of large-scale
 magnetic field in disks is still unknown. 
Large-scale magnetic fields in disks may be supplied externally and may also 
be generated through a kinematic alpha-omega dynamo (Reyes-Ruiz \& Stepinski 1999; 
Torkelssson \& Brandenburg 1994; 
Rudiger, Elstner \& Stepinski 1995).
MRI can also produce a turbulent small-scale MHD dynamo (Brandenburg et al. 1995; 
Hawley et al. 2001), 
however, the generation of large-scale magnetic field by MRI 
is not known to occur.
Based on the observational evidence of magnetic field 
 and the modeling of self-generated large-scale magnetic field (MHD dynamo) in disks,
we examine current-driven instabilities as a candidate for angular momentum
transport in an MRI stable magnetized disk with large currents.

In this paper, we explore the possibility of angular momentum transport 
in disks by current-driven instabilities in situations 
where MRI is stable (due to strong magnetic field)
and can not provide the required effective viscosity.  
We recognize three classes of situations that current-driven instabilities may be
important in transporting angular momentum in astrophysical disks; 1) 
in the inner disk region around a young magnetized protostar 2) in the innermost and active 
regions of weakly-ionized  protostellar and protoplanetary disks 3) in the upper and lower surfaces layers of a protoplanetary disk and magnetized coronas. Below we further
 discuss these situations.

The strong   
and complex large-scale magnetic fields detected in 
protostellar disks 
can provide the free energy (current) 
for current-driven instability and magnetic reconnection 
in these systems. 
Disk-star interaction and accreting matter to the stars along the complex magnetic 
field lines have been studied through MHD simulations by Long, 
Romanova \& Lovelace (2008) (and the reference therein). 
However, with open and complex field lines attached 
to the disk observed at radii only up to several stellar radii and with disk differential rotation, 
magnetic reconnection may be 
important in the inner region of a disk close to 
a young protostar (such as cTTs) and cause radial angular momentum transport.

Current-driven instabilities may also be important in protoplanetary and 
protostellar disks (e.g.
protosolar nebulas, disks around CTTs). 
However, in weakly ionized protoplanetary disks, the magnetic field 
might not be well coupled to the gas throughout the whole disk. 
Therefore, magnetic reconnection
can only play a role in some parts of these disks which are thermally 
ionized  (the innermost region r$<$0.1AU) or
nonthermally ionized  by cosmic rays (in the outer active region).
Other possible candidates for momentum transport in
disks are dynamo modes and flow-driven MRI. 
The notion of angular momentum transport through magnetic stresses
 from the 
oscillating dynamo modes has been examined by 
Stepinski \& Levy (1990) in a protostellar nebula in the absence of flow-driven 
MRI turbulence. It was 
estimated that a magnetic field strength of $10^2-10^3$G is needed 
for transport of angular momentum from protostar into a surrounding disk. 
Torkelssson \& Brandenburg (1994) 
also found stronger magnetic torques from quadropolar dynamo solutions. 
The effectiveness of MRI in transporting angular momentum transport in weakly-ionized
disks is limited and strongly depends on the ionization factor.
It has been shown that 
in the so-called
dead zone between the active layers in protoplanetary disks, MRI can not operate
(Gammie 1996; Fleming \& Stone 2003; Zhu, Hartmann \& Gammie 2010).~\nocite{gammie96,gammie10} However, 
including nonideal effects 
such as Hall effect and 
ambipolar diffusion have shown to change the stability and the saturation of 
MRI in weakly-ionized protoplanetary disks 
(Wardle 1999; Balbus \& Terquem 2001; Wardle 2007).~\nocite{wardle,balbus2fl}

In strongly magnetized coronas where MRI is stable, the role of current-driven 
instabilities can become significant. Magnetic activity at the surface of 
the disks, possibly originated from the internal dynamos, may produce 
coronas, which can transport angular momentum vertically and radially. 
Using local simulations, Miller \& Stone (2000) 
showed the generation of strongly magnetized corona (force-free) through 
an MRI-driven turbulence originated in the core of a 
disk with initially weak magnetic field. They also investigated
 angular momentum transport 
from MRI in a vertically extended local domain.
However, the possibility of angular momentum transport in the
strongly magnetized, MRI stable force-free
corona was not studied. Theoretical models have also been 
developed to study the possibility of angular momentum transport by 
the coronal magnetic field (Goodman 2003; Uzdensky \& Goodman 2008).
The current-driven instabilities can also become important 
in corona, upper and 
lower layers of 
protoplanetary disks, with significant currents, where MRI neither operates
in the core region (dead zone) nor is unstable in the coronas.

In this paper, we specify the 
equilibrium magnetic field (and corresponding current density) 
in disk geometry, which is chosen to be sufficiently 
strong that the flow-driven MRI is stable. We then 
study current-driven instabilities.
Unlike flow-driven MRI and pressure-driven interchange instabilities, 
which can be treated locally, current-driven instabilities are driven 
due to equilibrium gradients with scale length comparable to the 
global scale lengths.  
It is shown that gradients in global equilibrium parallel current provide
the driving source for both ideal and resistive 
current-driven (tearing) instabilities.  
Ideal current-driven 
instabilities are global modes (with global velocity perturbations) with
fast Alfvenic growth rates which are independent of resistivity. 
However, for tearing modes,  resistivity is important in the 
vicinity of the radius at which the wave number parallel to 
the magnetic field vanishes (i.e., $k_{\|}$ = k $\cdot$ B/B = 0). 
 Magnetic reconnection (or tearing) occurs at such radial locations. 
Although the reconnection occurs locally in radius, the tearing
 modes are global in their radial extent. 
 Therefore, tearing modes are  usually treated 
using asymptotic matching where solutions 
outside the reconnecting layer (outer layer or ideal $\eta=0$ solutions) 
are matched with the inner layer (reconnecting layer $\eta \neq 0$) solutions.

We investigate the physics of momentum transport 
through three sets of calculations of increasing 
completeness.  First, we perform linear stability 
analysis to establish conditions for instability 
and to investigate radial structure of the modes.
Second, through quasilinear analytic theory, we calculate
 the Maxwell stress in the outer ideal region and show its dependence on the 
azimuthal flow shear and instability growth rate.  We find that in the presence 
of azimuthal Keplerian flow, the radial and azimuthal magnetic fluctuations
are in phase and causes a nonzero Maxwell stress 
for a single (a single spatial Fourier component) current-driven mode and thus momentum transport. 
We also examine the structure of the stress 
for a single mode in the 
quasilinear regime. We show that the stress and 
the resulting momentum transport is localized to the vicinity 
of the reconnection layer.  However, an ideal 
current-driven instability has a global Maxwell 
stress, and therefore produces more global transport. 

Third, we compute the simultaneous nonlinear 
evolution of multiple modes, each corresponding to 
different reconnection layers (separated radially). 
With the inclusion of nonlinear mode coupling, the 
Maxwell stress becomes global, extending over the 
full plasma cross-section.  We perform nonlinear 
computations which are 1) weakly nonlinear driven 2) 
strongly nonlinear driven. In the weakly nonlinear driven case,
 many current-driven modes 
are linearly unstable while the effect of nonlinear coupling to stable modes is weak.
 In the strongly nonlinear case however, many stable modes are nonlinearly driven by
linearly unstable modes.
It is shown that in the weakly 
nonlinear driven case, 
the nonlinear structure of Maxwell stress is global 
mainly due to the growth of an ideal current-driven instability. 
However, in the strongly nonlinear case,  the global structure of the Maxwell 
stress is mainly due to the nonlinear coupling of many tearing modes.
We also estimate the effectiveness of the 
current-driven instabilities for angular momentum transport using the standard 
Shakura-Sunyaev $\alpha$ model for both the weakly nonlinear driven and strongly 
nonlinear driven cases. We find that Maxwell stress is much larger for the 
strongly nonlinear case compared to the weakly nonlinear case and  
 $\alpha_{SS}$ are about $10^{-3}$ and $10^{-2}$ for the weakly nonlinear driven and
strongly 
nonlinear driven case, respectively. 
 
The paper is organized as follows. The characteristics of momentum transport by current-driven 
instabilities and the model used are described in sections~\ref{sec:physics} and 
\ref{sec:model}, respectively.
Single mode calculations are presented in section
\ref{sec:single}. In Section \ref{sec:single1}, equilibrium 
and linear stability analysis are described. Analytical quasilinear calculation 
of Maxwell stress and the structure of single mode in quasilinear regime 
are presented in section \ref{sec:analy}.
We present full nonlinear multiple mode computations
 in section \ref{sec:nonlin}. Multiple computations for a 
weakly  nonlinear driven 
and a strongly nonlinear
driven case are 
presented in sections \ref{sec:nonlin1} and \ref{sec:nonlin2}, respectively.
We summarize in section \ref{sec:summ}.
 
\section{{\Large\textbf{Physical attributes of momentum transport by current-driven instabilities}}}
\label{sec:physics}

The momentum transport caused by current-driven instability has the same
origin in fluctuation-induced stresses as does the well-studied flow-driven MRI. 
In that sense, the mechanism for transport is the same, although the
underlying MHD instability is different in its energy source and spatial structure.
The change in flow is given by the MHD momentum equation
\begin{equation}
\rho \frac {\partial \textbf V }{ \partial t } =
- \rho \textbf V . \nabla\textbf V + \textbf J \times
\textbf B 
\end{equation}
For a rotating, cylindrical plasma we decompose the flow and magnetic fields into
spatially mean and fluctuating quantities, where mean values (denoted by $<>$)
are averaged over axial and azimuthal directions and fluctuating values subtract
out the mean. If we then average the above equation over the axial and azimuthal
directions we find
\begin{equation}
\rho \frac {\partial <\textbf V> }{ \partial t } =
- \rho <\textbf V . \nabla\textbf V> + <\textbf J \times
\textbf B> 
\label{eq:mom}
\end{equation}
which describes the time evolution of the mean flow which depends upon radius
and time. We see that the mean flow can evolve from fluctuating flow (arising
from a Reynolds stress, the first term on the RHS) and fluctuating magnetic field
(the Maxwell stress or Lorentz force of the second term on the RHS). These terms,
quadratic in the fluctuations, can be nonzero upon spatial averaging.
MHD current-driven instabilities produce spatially fluctuating quantities
that, through the Reynolds and Maxwell stress alter the radial profile of the axial
and azimuthal mean flow. These forces can decrease the flow at one radius, while
increasing it another, essentially transporting momentum radially. In this sense, the
mechanism is identical to momentum transport from the well-studied magneto-
rotational instability. Crucial differences lie in the energy source, spatial structure,
and dynamics of the underlying instabilities.

The MRI is a global ideal instability (occurs at zero resistivity) with the
stresses acting globally in radius. The current-driven tearing instability also has an
amplitude that is global in radius, but the forces on the RHS of the Eq.~\ref{eq:mom}
act locally in radius for a single instability. For a single tearing instability 
 magnetic reconnection occurs at one radial location that for which
the wavelength parallel to the equilibrium magnetic field becomes infinite (Biskamp 1993). That is,
the instability amplitude is constant along the helical magnetic field at this radius,
although it varies in the azimuthal and axial directions. Only the vicinity of this
radius is electrical resistivity important. The resistive layer of narrow radial extent,
within which reconnection occurs, functions as a boundary layer between two ideal
regions. Within this region amplitudes of velocity and current density can become
large, leading to locally strong forces that alter flow.

An additional difference between the MRI and tearing instability is that
momentum transport in the MRI can be understood as the instability acting to
diminish its energy source (flow gradient). For the tearing instability the analogue
is the reduction of the current density, which is a strong effect of the instability.
Reduction of the flow gradient can be viewed as a parasitic effect of the tearing
mode. Interestingly, the MRI also alters the magnetic field (Ebrahimi, Prager \& Schnack 2009), so both instabilities alter the flow and field.
We should note that an ideal current-driven instability 
despite having the same energy source as tearing instability, exhibits different
characteristics.
A major difference is that for an ideal current-driven instability (with no reconnecting layer
in the plasma volume), the fluctuation-induced forces are global as will be shown below.

The tearing mode, being a resistive instability, does not lend itself to simple
analytic calculation of instability-induced transport. 
However, in section~\ref{sec:analy}, an
analytic calculation of the transport is provided for the ideal region.
We find that 
in the presence of azimuthal flow, the radial and azimuthal magnetic fluctuations are in phase and 
results in a non-zero Maxwell stress. The resulting stress (and the 
direction of angular momentum transport) depends on the 
global equilibrium as well as the global radial mode structure.
We therefore employ quasilinear computations to show the structure of 
stresses and outward momentum transport.
It will be shown that the structure of the stresses for ideal 
current-driven and tearing modes are very different, therefore they can affect the 
momentum transport differently. With multiple tearing modes
present, nonlinear three-wave interactions alter the structure of the modes and the
forces broaden radially.

\section{{\Large\textbf{The model}}}
\label{sec:model}
\begin{figure} 
\includegraphics[]{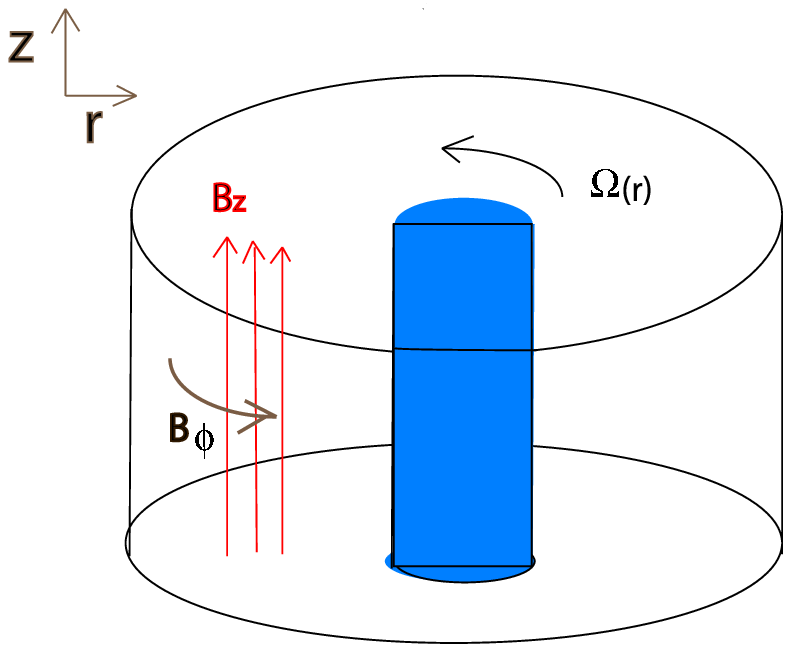}
\caption{A current-carrying rotating disk configuration.}   
\label{fig:fig1} 
\end{figure}

Throughout this work, we employ the MHD equations in doubly 
periodic $(r,\phi,z)$ cylindrical geometry for both the analytic
and computational studies.  
All variables are decomposed as 
$f(r,\phi,z,t)=\sum_{(m,k)} \widehat f_{m,k}(r,t) e^{i(m \phi + kz)}=<f(r,t)>+\widetilde f(r,\phi\,z,t)$, 
where $<f>$ is the \textit {mean} $(m=k=0)$ component, and $\widetilde f$ 
is the fluctuating component 
(i.e., all other Fourier components with $m \ne 0$ and $k \ne 0$).  
Note that the mean component is a function of radius.  
We consider an azimuthal equilibrium flow 
$ \textbf V_0 = V_{\phi}(r) \hat{\phi}$ in a current-carrying disk configuration plasma. 
(Here, $V_{\phi}(r)$ is the equilibrium value of the mean azimuthal flow $< V_{\phi}>$).
In order to excite the current-driven instabilities, 
both vertical magnetic field and the azimuthal
magnetic field, $\textbf B = B_z(r) \hat{z} + B_{\phi}(r) \hat{\phi}$, 
are imposed (Fig.~\ref{fig:fig1}).  The single fluid MHD equations are,  
\begin{eqnarray}
\frac {\partial \textbf A }{ \partial t } &=& -\textbf{E} = 
S\textbf V\times \textbf B - \eta \textbf J\\
\rho \frac {\partial \textbf V }{ \partial t } &=&
-S \rho \textbf V . \nabla\textbf V + S\textbf J \times
\textbf B +P_m \nabla^2 \textbf V -S \frac{\beta_0}{2}\nabla P -S \rho \nabla \Phi \\
\frac {\partial P }{ \partial t } &=&
-S\nabla \cdot (P \textbf V) - S (\Gamma -1) P \nabla \cdot \textbf V\\
\frac {\partial \rho }{ \partial t } &=&
-S\nabla \cdot (\rho \textbf V) \\
\textbf B &=& \nabla \times \textbf A\\
\textbf J &=& \nabla \times \textbf B
\end{eqnarray}
where the variables, $\rho, P, V, B, J, $, $\Gamma$, and $\Phi$ are 
the density, pressure, velocity, magnetic field, current, 
ratio of the specific heats, and gravitational potential respectively.
Time and radius are normalized to the resistive diffusion time 
$\tau_R = 4 \pi {r_2}^2/c^2\eta_0$ and the outer radius $r_2$, 
making the normalized outer radius unity. 
Velocity \textbf{V} and 
magnetic field \textbf{B} are normalized to the Alfv\'en velocity
 $V_A$, and the magnetic field on axis $B_0$, respectively.
The parameter $S = \frac {\tau_R}{\tau_A}$ 
is the Lundquist
number (where $\tau_A = r_2/V_A$), and $P_m= \frac {\tau_R}{\tau_{vis}}$  
 measures the ratio of 
characteristic viscosity $\nu_0$ to resistivity $\eta_0$ (the magnetic Prandtl number), 
where $\tau_{vis}= 4 \pi {r_2}^2/c^2\nu_0$
is the  viscous 
diffusion time.
The factor $\beta_0 =8\pi P_0/B_0^2$ is the beta normalized 
to the axis value. The resistivity 
profile $\eta$ is uniform.  The boundary conditions in 
the radial direction are as is appropriate to dissipative 
MHD with a perfectly conducting boundary: the tangential electric 
field, the normal component of the magnetic field, and the normal 
component of the velocity vanish, and the tangential component of 
the velocity is the rotational velocity of the wall.  
The azimuthal $(\phi)$ and axial $(z)$ directions are periodic.

We pose an initial value problem that consists of the equilibrium 
plus a perturbation of the form 
$\tilde f(r, \phi,z,t)=\tilde f_{m,k} (r,t) \mathrm exp( im\phi +ikz)$. 
Equations (1-6) are then integrated forward in time using the DEBS code.   
The DEBS code uses a 
finite difference method with a staggered grid for radial 
discretization and pseudospectral method for 
azimuthal and vertical coordinates.  
In this decomposition, each mode satisfies a separate equation of the 
form $\partial \tilde f_{m,k} / \partial t = L_{m,k}  \tilde f_{m,k} +\sum_{(m^{'},k^{'})} N_{m,k,m^{'},k^{'}}$, where $L_{m,k}$ is a 
linear operator that depends on $\tilde f_{0,0}(r,t)$, 
and $N_{m,k,m^{'},k^{'}}$ is a nonlinear term that 
represents the coupling of the mode $(m,k)$ to all other modes $(m^{'},k^{'})$.  (
This latter term is evaluated pseudospectrally.) 
The time advance is a combination of the leapfrog 
and semi-implicit methods (Schnack et al. 1987).    

We present three types of initial value computations, linear, nonlinear single mode and
fully nonlinear computations. In linear computations, 
the initial conditions consist of an equilibrium $<f(r)>$ plus 
a single mode $\tilde f_{m,k}(r,0) \mathrm exp( im\phi +ikz)$ perturbations. 
Only the mode $(m,k)$ is then evolved; in particular,  
the equilibrium (the $m=0$, $k=0$ mode, or $<f>$) is \textit{not} 
evolved, and remains fixed in time.  In the nonlinear single mode computations, 
however,  the $m=0$, $k=0$ component (the background) is allowed to evolve self-consistently.
The evolution of the background profile $\tilde f_{0,0}$ 
 can affect the evolution of the mode $(m,k)$ and cause the mode to saturates. 
In a fully nonlinear computation, all modes are initialized 
with small random amplitude and are  evolved in time, 
including the full nonlinear term ($N_{m,k,m^{'},k^{'}}$).

The plasma rotates azimuthally with a mean Keplerian flow $<V_{\phi}(r)> \propto r^{-1/2} $.
The initial (i.e., at $t=0$)  radial \textit {equilibrium} 
force balance (Eq. 2) is satisfied by $ \frac{\beta_0}{2}\nabla p + \rho \nabla \Phi = \rho V_{\phi}^2/r$, where $\nabla \Phi =GM/r^2$, 
and a magnetic force free condition $\textbf{J}\times \textbf{B} =0$. 
The initial pressure and density profiles are assumed to be radially uniform.
Pressure and density are evolved; however, they remain fairly uniform during the 
computations.

We consider a cylindrical disk-shaped plasma with aspect ratio 
 L/($r_2 -r_1$) 
(where L is the vertical height, and $r_1$ and $r_2$ 
are the inner and outer radii, respectively)(Fig.~\ref{fig:fig1}). 
The inner and outer radial boundaries are perfectly conducting, concentric 
cylinders that can rotate independently at specified rates.
Periodic boundary conditions are used 
in the vertical and azimuthal directions.

The aspect ratio (L/($r_2 -r_1$))
used in the nonlinear computations is 1.3. The nonlinear computations are performed
in a thick-disk approximation where vertical and radial 
distances are of the same order.
 The range of parameters used in the computations 
is $\beta= 1 - 10$, $S=10^4$, $P_m=0.1-20$.

\section{{\Large\textbf{Single mode calculations }}}
Here we present linear and quasilinear single mode computations 
for current-driven instabilities. We first choose equilibria 
which are unstable for current-driven instabilities and then perform
linear computations (section~\ref{sec:single1}). To study whether 
a single current-driven instability transports momentum, we examine
the Maxwell stress in the ideal MHD region (outer region) through
quasilinear analytical calculations (section~\ref{sec:analy}).
We also examine the structure of quasilinear stresses for a single mode
 (with specific m and n) with the equilibrium 
described in section~\ref{sec:single1}.

Unlike flow-driven MRI and pressure-driven interchange instabilities, 
which can be treated locally, current-driven instabilities have 
global characteristics. Gradients in global equilibrium parallel current provide
the driving source for the ideal and resistive current-driven instabilities.  
Ideal current-driven 
instabilities, so-called kink modes, are helical long wavelength structures with global
(broad radial extent) velocity perturbations. These ideal modes do not scale with 
resistivity and have fast Alfvenic growth rates.  In an ideal plasma, 
the restoring force is infinite and fluid is frozen to the field. However, 
in the presence of resistivity, the field lines can break up
and at locations where the parallel wave number is zero, (so called resistive or reconnecting layer)  
field lines can reconnect. Both structure and growth rate of a 
resistive current-driven mode (tearing mode) 
are affected by resistivity. The solutions around the reconnecting layer are solved by including 
the resistivity (so called inner layer solutions) and are asymptotically matched with solutions outside the reconnecting layer (so called outer layer solutions). 
We will show that the structure of the stresses for ideal 
current-driven and tearing modes are very different, therefore they can affect the 
momentum transport differently.    

\label{sec:single}

\subsection{{\Large\textbf{Linear computations}}}
\label{sec:single1}

\begin{figure}
\includegraphics[]{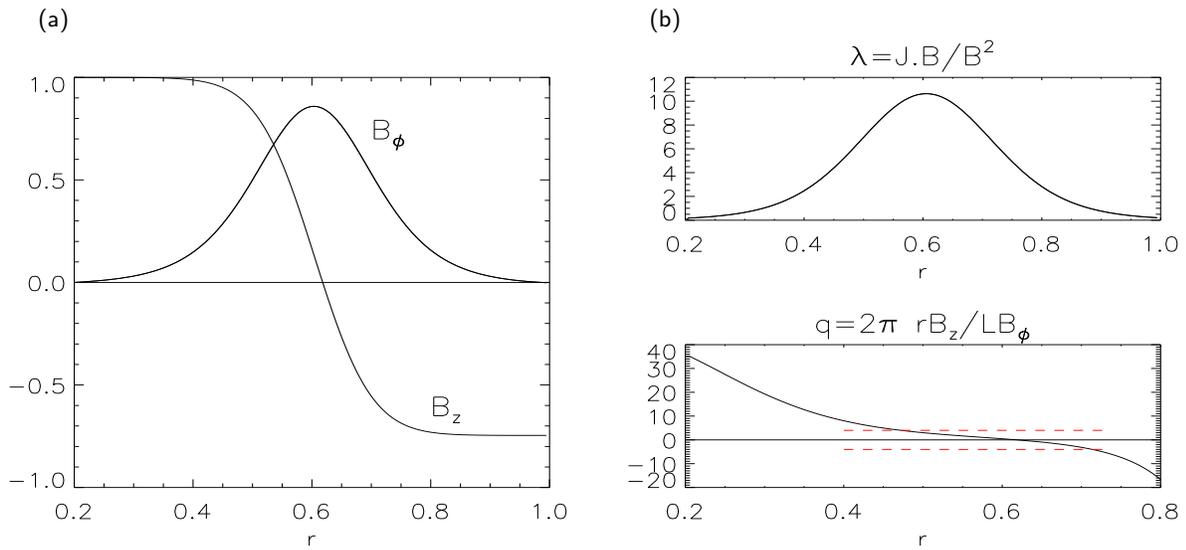}
\caption{Radial profiles of (a) equilibrium azimuthal and 
vertical magnetic fields (b) 
$\lambda=\frac{J_{||}}{B}$ and axial winding number q= $2 \pi r B_z  /LB_{\phi}$.}
\label{fig:fig2}
\end{figure}

We first choose equilibrium profiles that are unstable for current-driven tearing instabilities.
We start with a force free plasma $\textbf{J}\times \textbf{B} =0$. Current 
flows parallel to the magnetic field line $\textbf J_{||}=\lambda (r) \textbf{B}$, where $
J_{||}$ is the parallel current.
 Figure ~\ref{fig:fig2}(a) shows a typical equilibrium 
$\lambda (r)= \textbf J_{||}/\textbf{B}$ profile used
in the computations at t=0. The equilibrium vertical and azimuthal 
magnetic field components 
can be obtained from the force balance equation 
$\nabla \times \textbf{B} = \lambda(r) \textbf{B}$, and are shown 
 in Fig.~\ref{fig:fig2}(a). The magnitude of the magnetic fields is large
enough to make this equilibrium MRI stable.

As described in the previous section, perturbations are assumed in the form of
$\mathrm{exp(i k \textbf r)} \sim \mathrm{exp (im\phi + i 2 \pi n z/L)}$ in cylindrical geometry 
, where m and n are the 
azimuthal and axial mode number respectively ($k_z=2\pi n/L$). 
At locations where parallel wave number 
is zero $\textbf{k} \cdot \textbf{B} = m B_{\phi}/r + 2 \pi n B_z/L=0$ 
(so called resonant or reconnecting surfaces),
 resistivity becomes important and 
reconnection can occur. 
From this resonant condition, a field line winding number $q = -m/n= 2 \pi rB_z/LB_{\phi}$
can be defined. The profile of axial winding number q is shown in Fig.~\ref{fig:fig2}. 
 Here, 
conventionally positive q corresponds to a negative axial mode (n) and vice versa. 
It will be shown that depending on the equilibrium profile ($\lambda$), 
 current-driven modes with $-5\leq q \leq 5 $ can be linearly unstable. 

To verify that the equilibrium is tearing mode unstable, 
we first perform linear single mode computations (single m and n) 
%with the equilibrium shown in Fig.~\ref{fig:equil},  
in the absence of azimuthal flow. The radial mesh points used is 250. 
The modal structure and the growth rates of the tearing modes depends on the 
equilibrium profile, i.e parallel current profile $\lambda(r)$. Therefore, 
we perform four set of linear computations with four different equilibria. 
The four equilibria
 are distinguished by the maximum value of the equilibrium $\lambda=J_{||}/B$ profile.
We find that both resistive current-driven modes (tearing modes) and 
ideal current-driven modes with azimuthal 
modes numbers m=0,1,2,3 are linearly unstable for sets of equilibria 
with $\lambda_{\mathrm{max}}$
=10.6 (shown in Fig.~\ref{fig:fig2}), $\lambda_{\mathrm{max}}$ =14.2, and
 $\lambda_{\mathrm{max}}$ =17.2. For equilibria with $\lambda_{\mathrm{max}}$ =9, only 
m=0,1,2 tearing modes are linearly unstable. Below, we further discuss the 
mode structure and the
properties of tearing modes and ideal current-driven modes.

\begin{table}
\caption{Growth rates ($\gamma \tau_A$) of tearing modes (m,n), for two $\lambda_{max}$. }
\begin{tabular}{|c|c|c|c|c|c|c|c|r|}\hline
$\lambda_{max}$&(1,0)&(1,1)&(1,-1)&(0,1)&(2,1)&(2,-1)&(3,1)&(3,-1)\\ 
\hline
9&stable&0.030&0.031&0.0362&0.0155&0.0165&stable&stable\\
\hline
10.63&0.058&0.059&0.045&0.057&0.048&0.0445&0.029&stable\\
\hline
14.2&0.23&0.148&0.101&0.132&0.141&0.0557&0.116&0.0105\\ 
\hline
17.2&0.3&0.31&0.21&0.275&0.3&0.14&0.255 &0.082\\ 
\hline
%\caption{Linear growth rates for two cases}
\end{tabular}
\end{table} 
 
\begin{figure}
\includegraphics[]{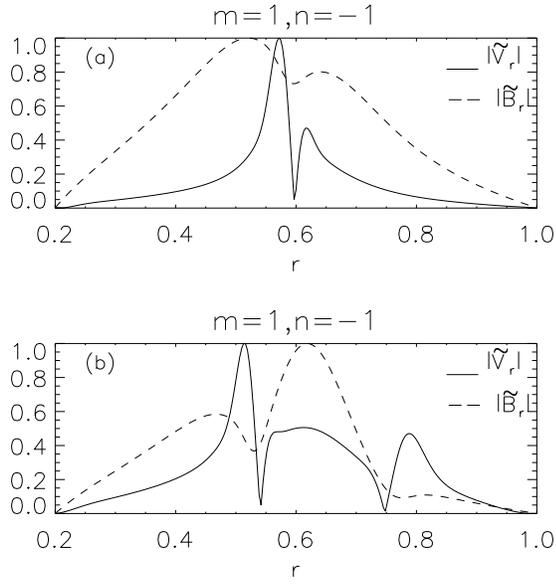}
\caption{Radial structure of linear magnetic and velocity 
eigenfunctions for m=1 unstable tearing modes, $S=10^4$, $P_m=1$ (a) $\lambda_{max}$ =10.6 (b) $\lambda_{max}$ =17.2.} 
\label{fig:fig3}     
\end{figure}

In the presence of resistivity, current-driven 
instabilities (tearing) can become linearly unstable.
The radial structure of 
radial magnetic and velocity  eigenfunctions for m=1 tearing 
mode with axial modes number (n=-1) are shown in Fig.~\ref{fig:fig3}(a)
(for $\lambda_{\mathrm{max}}=10.6$) and Fig.~\ref{fig:fig3}(b) (for 
$\lambda_{\mathrm{max}}=17.2$).
As it is seen, the reconnecting component of magnetic field ($\widetilde B_r$) 
is nonzero around the reconnecting surface 
[r=0.6 for (1,-1)in Fig.~\ref{fig:fig3}(a)].
Moreover, the radial fluctuating velocity is concentrated and change sign around 
the reconnecting surface.
One of the characteristics of tearing instability is the jump in the 
logarithmic derivative of $\widetilde B_r$ across the resistive layer ($\Delta^{\prime} =
(\widetilde B_r^{\prime}|_{rs}^+ - \widetilde B_r^{\prime}|_{rs}^-)/\widetilde B_r|_{rs}$), where rs denotes reconnecting surface.. For tearing mode
to be unstable, this jump should be positive (Furth et al. 1963).~\nocite{fkr}
The growth rate of the tearing mode 
scales with resistivity and $\Delta^{\prime}$ as $\gamma_{\mathrm{tearing}} \propto \eta^{3/5} \Delta^{\prime 4/5}$ (or $\propto S^{-3/5} \Delta^{\prime 4/5}$). 
As is seen in Fig.~\ref{fig:fig3}, $\Delta^{\prime}$
 is positive for m=1 tearing modes, and is larger in Fig.~\ref{fig:fig3}(b) with
larger free energy (current gradient $\lambda_{\mathrm{max}}$), which results in larger
growth rate. We also note that for larger $\lambda_{\mathrm{max}}$, (1,-1) mode becomes a double tearing mode (reconnects at two locations), and radial velocity perturbation changes
sign twice in radius (absolute values of the eigenfunctions are shown in Fig.~\ref{fig:fig3}(b)).

\begin{figure}
\includegraphics[]{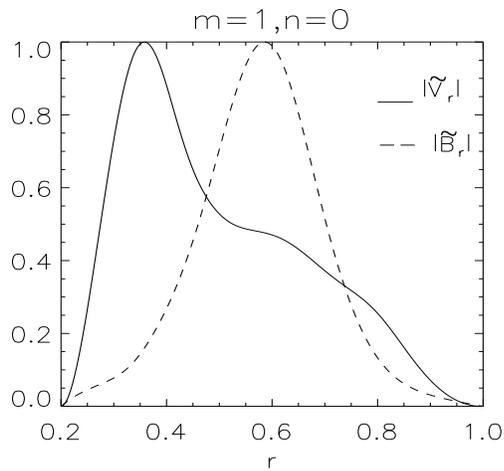}
\caption{Radial structure of linear magnetic and velocity eigenfunctions for the ideal
m=1, n=0 current-driven mode.}
\label{fig:fig4}
\end{figure}

The growth rates of m=0-3 tearing modes for four set of  $\lambda$ profiles
 with $\lambda_{\mathrm{max}}$=9, 10.6, 14.2, and 17.2 are given in Table 1.   
For smaller $\lambda_{\mathrm{max}}$, the tearing mode growth rates are smaller, and
the growth rates increase with $\lambda_{\mathrm{max}}$ approaching 
the ideally unstable limit (approaching Alfvenic growth rates). 
Modes with azimuthal mode
numbers 0-3 and axial mode numbers 1,-1 have tearing mode structure 
with a corresponding reconnecting surface. However, mode with axial mode number zero (1,0), is ideally
unstable for equilibria with $\lambda_{\mathrm{max}}$=10.6, 14.2, and 17.2.
The radial structure of this mode is shown in 
Fig.~\ref{fig:fig4}
which has a kink-like characteristic.
Unlike the tearing modes, ideal current-driven modes does 
not necessarily have to have  a 
reconnecting surface (with $\textbf{k} \cdot \textbf{B}$) 
within the plasma  and the fluctuating velocity components 
can be very global without changing sign (kink-like). The growth rates
of the ideal current-driven modes also don't scale with resistivity.
 It has been shown that
 the current-driven modes are unstable for the equilibria chosen
(Fig.~\ref{fig:fig2}) and therefore, we can
 study the effect of both tearing and 
ideal current-driven modes on the momentum transport through
nonlinear computations (section \ref{sec:nonlin}).

\subsection{{\Large\textbf{Analytical quasilinear calculations}}}
\label{sec:analy}
In order to investigate momentum transport from current-driven 
instabilities, we obtain MHD stresses using quasilinear calculations of a single mode
with an initial azimuthal Keplerian flow. The question is whether a single
 current-driven mode can transport momentum and affect the azimuthal flow profile.
To obtain more insight into this question, we first analytically examine the 
ideal MHD equations with azimuthal
flow in cylindrical geometry. We then construct the Maxwell stress 
$<r^2 \widetilde B_{\phi} \widetilde B_r> = 2 Re( r^2 \widetilde B_{\phi}^*\widetilde B_r)$ from the linearized solutions in the outer (ideal) region. 
Through analytical quasilinear calculations, we aim to identify 
 whether the Maxwell stress is nonzero for a current-driven instability and therefore 
momentum can be transported. We do not intend to analytically solve the complete sets 
of solutions (inner and outer solutions), and we only present the simplified 
outer solutions. To construct the radial structure of the quasilinear stresses, 
we use the eigenfunctions from the linearized 
computations.

The linearized incompressible ideal MHD equations 
in the presence of mean azimuthal flow are,
\begin{equation}
\rho (\frac{\partial \widetilde{\textbf{V}}}{\partial t}+ (\textbf V_0 \cdot \nabla) \widetilde{\textbf{V}} + (\widetilde{\textbf{V}} \cdot \nabla) \textbf V_0) = -\nabla( \textbf B_0 \cdot \widetilde{\textbf{B}})+(\textbf B_0 \cdot \nabla) \widetilde{\textbf{B}}+(\widetilde{\textbf{B}} \cdot \nabla) \textbf B_0
\end{equation}
 \begin{equation}
\frac{\partial \widetilde{\textbf{B}}}{\partial t}= \nabla \times (\widetilde{\textbf{V}} \times \textbf B_0) +  \nabla \times (\textbf V_0 \times  \widetilde{\textbf{B}})
\end{equation}

We assume perturbation of the form Q(r, $\phi$,z,t)=Q(r) exp(i $\omega$ t + i (m $\phi$ + $k$ z)), and the equilibrium magnetic field and flow are 
$\textbf{B} = B_{\phi}(r) \textbf{e}_{\phi} + B_z(r) \textbf{e}_z$
 and $\textbf{V}_0 = V_{\phi}(r) \textbf{e}_{\phi}$ respectively.
The linearized equations can be combined and be presented in 
the form of one ordinary 
differential equation for $\widetilde{V}_r$,

\begin{equation}
\label{eq:vrout}
(r \widetilde{V}_r)^{\prime \prime}  + a_1 (r \widetilde{V}_r)^{\prime} + a_2 (r \widetilde{V}_r)=0
\end{equation}
where,
\begin{equation} 
\begin{split}
\label{eq:vrout2}
a1&= \frac{1}{A} \frac{dA}{dr} - \frac{m G}{r \bar{\omega} (1-M^2)} + \frac{2 m \rho \bar{\omega} V_{\phi}}{r^2 F^2 (1-M^2)}\\
a2&=\frac{1}{A}[\frac{d}{dr}(\frac{m G A }{r \bar{\omega} (1-M^2)}) - 
\frac{d}{dr}(\frac{2 m B_{\phi} F}{r \bar{\omega} (m^2 + k^2 r^2)}) \\&+ \frac{2 B_{\phi}}{r \bar{\omega}}
(\frac{B_{\phi}}{r})^{\prime} + \frac{4 k^2 B_{\phi}^2}{r \bar{\omega} (1-M^2) (m^2 + k^2 r^2) } 
-\frac{F^2 (1-M^2)}{r \bar{\omega}}\\&- (\frac{2 B_{\phi} m^2 F}{r^2 \bar{\omega}^2(m^2 + k^2 r^2)}) (V_{\phi}^{\prime} -\frac{V_{\phi}}{r}) - (\frac{4 k^2 F B_{\phi} M^2}{\bar{\omega}^2 (m^2 + k^2 r^2)(1-M^2)})\frac{V_{\phi}}{r} \\&- (\frac{4 \rho k^2B_{\phi}}{F(m^2 + k^2 r^2) (1-M^2)})\frac{V_{\phi}}{r} - \frac{2 \rho k^2 V_{\phi} G}{\bar{\omega} (m^2 + k^2 r^2) (1-M^2)}  ]
\end{split} 
\end{equation}
and;
\begin{eqnarray*} 
  A= \frac{r F^2 (1-M^2)}{\bar{\omega} (m^2+k^2 r^2)},&& G = (1-M^2)V_{\phi}^{\prime} 
- (1+M^2)V_{\phi}/r 
\end{eqnarray*} 
\begin{eqnarray*} 
\bar{\omega} = \omega +m V_{\phi}/r ,&& M = \sqrt{\rho} \frac{\bar{\omega}}{F}
\end{eqnarray*} 
\begin{eqnarray*} 
F= \frac{m B_{\phi}}{r}+ k B_z ,&& k_{\perp}B = - m B_z/r+ k B_{\phi}
\end{eqnarray*} 

where $\omega= \omega_r + i \gamma$, and $\omega_r$ and $\gamma$ are the 
real frequency and the  the growth rate, respectively. 
The equation for parallel magnetic field perturbations 
$\frac{\widetilde{\textbf{B}} \cdot \textbf{B}}{B} = \widetilde{B}_{||}$ 
in terms of radial velocity 
perturbation is obtained,

\begin{equation}
\label{eq:bbout} 
\widetilde{B}_{||}=i \frac{A}{B} [\frac{\partial}{\partial r}(r \widetilde{V}_r)-(\frac{2 m B_{\phi}}{(1-M^2)F r^2}+\frac{m G}{r \bar{\omega} (1-M^2)}) (r \widetilde{V}_r)]
\end{equation}

To obtain the Maxwell stress term, we write the parallel magnetic 
field perturbation Eq.~\ref{eq:bbout} in terms of radial magnetic field perturbation 
$\widetilde{B}_r = F \widetilde{V}_r/\bar{\omega}$ and 
  the parallel magnetic 
field perturbation  without 
mean flow $\widetilde{B}_{||}^{(0)}$,
\begin{equation}
\label{eq:bbout2} 
\widetilde{B}_{||}= (1-M^2) \widetilde{B}_{||}^{(0)}
- i M^2 (\frac{2 m B_{\phi}}{r B (m^2+k^2 r^2)})(r  \widetilde B_r)
+i M^2(\frac{2 m F V_{\phi}}{r \bar{\omega}  B (m^2+k^2 r^2)})(r  \widetilde B_r)
\end{equation}

where $\widetilde{B}_{||}^{(0)}$$ =  i  [F r 
(r \widetilde{B}_r)^{\prime} - r (k_{\perp}B) \lambda (r \widetilde{B}_r)] /(B (m^2 + k_z^2 r^2))$ is the 
parallel magnetic field in the 
absence of mean flow obtained  from Eq.\ref{eq:bbout}, 
and also known from Newcomb~\nocite{newcomb} equation (Newcomb 1960), 
and $\lambda=J_{||}/B$ . 
In the absence of mean flow, it can be seen that $\widetilde{B}_{||}^{(0)}$ 
is purely imaginary and $\widetilde{B}_r$ and $\widetilde{B}_{||}^{(0)}$ 
are out of phase,  
and therefore parallel and azimuthal Maxwell stresses vanish for the ideal region.
However, with the azimuthal mean flow the Maxwell stress 
$<r^2 \widetilde B_{\phi} \widetilde B_r> = 2 Re( r^2 \widetilde B_{\phi}^*\widetilde B_r)$ from a current-driven 
mode is nonzero and 
is obtained,
\begin{equation} 
\begin{split}
\label{eq:maxwell}
<r^2 \widetilde B_{\phi} \widetilde B_r>= 2 \gamma
\left(2 k \rho r^2 \frac{(\omega_r +m V_{\phi}/r)}{k_{\perp} F^2}   |
\widetilde{B}_{||}^{(0)}| \widetilde{B}_r\right) \\+ 2 
\gamma
\left(4 k  \rho r^2\frac{(\omega_r +m V_{\phi}/r)}{ k_{\perp}  (m^2 + k^2 r^2)F^2 B} 
 m B_{\phi} \widetilde{B}_r^2 \right)\\- 2 
\gamma
\left(2 k  \rho r^2\frac{m V_{\phi}}{ k_{\perp}  (m^2 + k^2 r^2)F B} 
  \widetilde{B}_r^2 \right)
\end{split}
\end{equation}
where,
\begin{equation}
\label{eq:bphi} 
\widetilde{B}_{\phi}=\frac{k \widetilde{B}_{||}}{k_{\perp}} - \frac{i B_{z}}{(Brk_{\perp})}\frac{\partial}{\partial r}(r \widetilde{B}_r)
\end{equation}
The azimuthal Maxwell stress Eq.~\ref{eq:maxwell} has been calculated using Eqs.~\ref{eq:bbout2},
and~\ref{eq:bphi}, assuming $\widetilde B_r$ is purely real.
 With mean azimuthal flow,
 Eq.~\ref{eq:maxwell} shows that the joint effect of mode growth 
and mean flow produces a nonzero stress term in the outer region. 
We note that for an ideal current-driven mode without a reconnecting surface (F$\neq$ 0 everywhere), 
the stress term can be global. 
The sign and the structure of the resulting 
stress depend on the global equilibrium 
as well as the global radial mode structure and do not follow from 
Eq.~\ref{eq:maxwell} alone. 
However, for a special case of an ideal current-driven
mode m=1, n=0 (k=n/R=0), discussed in section~\ref{sec:single1},
 we can examine the sign of the Maxwell stress 
using local WKB approximation.
For this mode the Maxwell stress is simplified as
 $<r^2 \widetilde B_{\phi} \widetilde B_r> = 2 Re\{[\frac{i r^2}{m}\frac{\partial}{\partial r}(r \widetilde{B}_r)]^*\widetilde{B}_{r}\}$, where $\widetilde B_r$ is complex, and we have used 
Eq.~\ref{eq:bphi} for the toroidal magnetic perturbation $\widetilde B_{\phi}$ 
(or equivalently from $ \nabla \cdot \textbf B =0$).
Using a local approximation, a negative Maxwell stress 
$<r^2 \widetilde B_{\phi} \widetilde B_r> = -2 r^3 \frac{k_r}{m} \widetilde B_r^2$ 
is obtained, where $k_r$ is the local radial wavenumber. The negative Maxwell stress results in 
a bi-directional Lorentz force and causes outward momentum transport.
 Although WKB is not a valid approximation 
for the current-driven instability and does not provide the structure of the transport, 
it demonstrates an outward transport, which is consistent with the global
solutions given below. 

\begin{figure}
\includegraphics[]{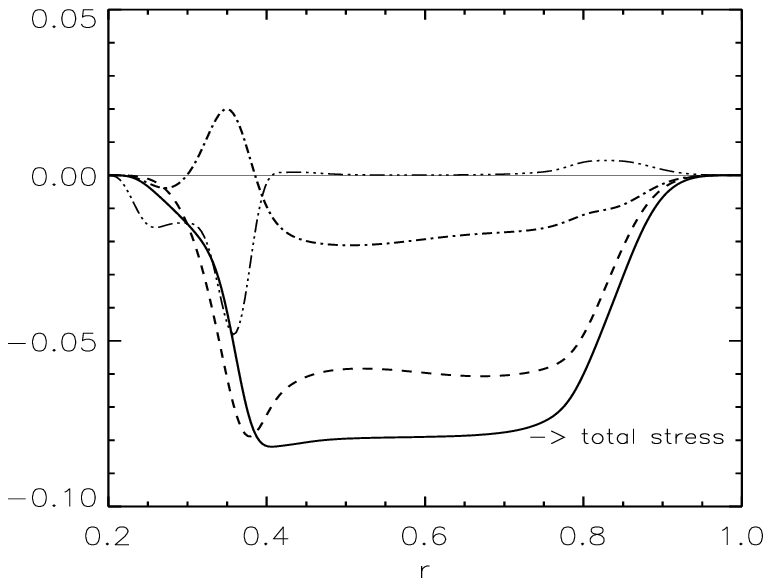}
\caption{Radial structure of quasilinear total stress (solid line), Maxwell stress
$<r^2 \widetilde B_{\phi} \widetilde B_r>$ (dashed line), Reynolds stress 
$-<r^2 \widetilde V_{\phi} \widetilde V_r>$ (dashed-dotted), and total azimuthal fluid 
force $<\widetilde{\textbf{J}} \times \widetilde{\textbf{B}}>_{\phi} - 
\rho <\widetilde{\textbf{V}} \cdot \nabla \widetilde{\textbf{V}}>_{\phi} $ 
(dashed-dotted-dotted) constructed from the linear eigenfunctions
for ideal current-driven mode m=1 n=0 mode.} 
\label{fig:fig5}
\end{figure}

To obtain the radial structure of stresses, we employ quasilinear computations. 
Using the linear eigenfunctions for $\widetilde B_r$ (shown in Fig.~\ref{fig:fig4}),
the radial structure of the Maxwell stress term for an ideal current-driven mode, m=1,n=0,
is shown in Fig.~\ref{fig:fig5}. 
Similarly, the quasilinear Reynolds stress term,
 $<r^2 \widetilde V_{\phi} \widetilde V_r>$, can also be constructed from the linearized solutions for the m=1, n=0 mode (shown in Fig.~\ref{fig:fig5} by dashed-dotted line).
We note that the total stress term, $<r^2 \widetilde B_{\phi} \widetilde B_r> - <r^2 \widetilde V_{\phi} \widetilde V_r>$, is negative everywhere, which causes
a transport of angular momentum outward. 
Total azimuthal fluid force consisting  of Lorentz and inertia terms 
  $<\widetilde{\textbf{J}} \times \widetilde{\textbf{B}}>_{\phi} - 
\rho <\widetilde{\textbf{V}} \cdot \nabla \widetilde{\textbf{V}}>_{\phi} =   \frac{1}{r^2}[\frac{d}{d r}<r^2 \widetilde B_{\phi} \widetilde B_r> - \rho \frac{d}{d r}<r^2 \widetilde V_{\phi} \widetilde V_r>]$, is bi-directional and causes transport of momentum outward, 
as shown in Fig.~\ref{fig:fig5}.
 
\begin{figure}  
\includegraphics[]{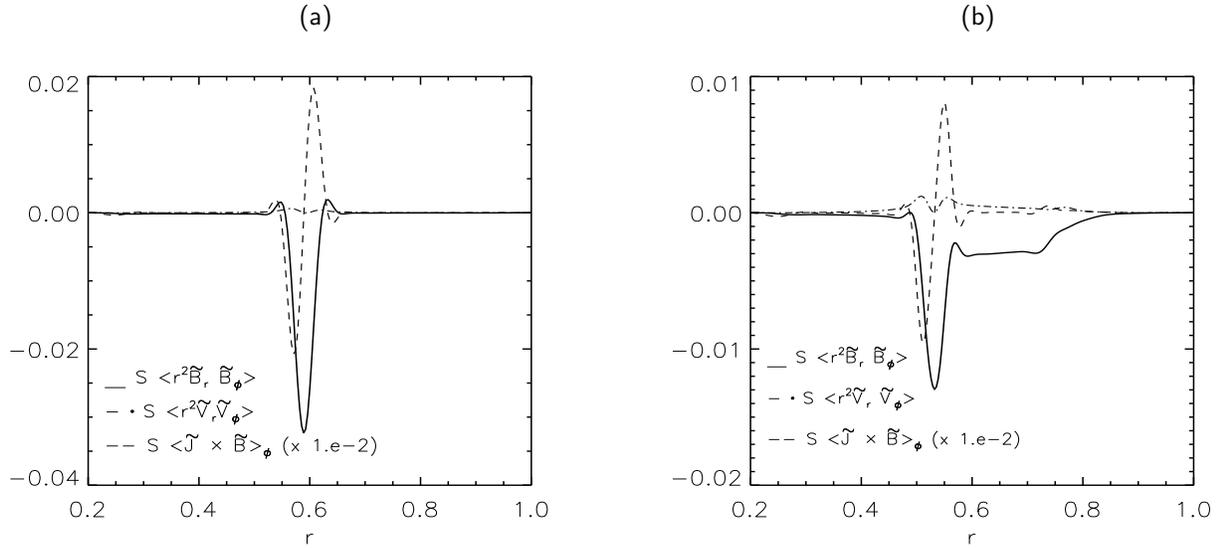}
\caption{Radial structure of Maxwell, 
Reynolds stresses and the Lorentz force during 
the linear phase before the mode saturation for m=1, n=-1 (a) single tearing mode 
with $\lambda_{max}=10$ and (b) a double tearing mode with
 $\lambda_{max}=17$, $S=10^4$, $P_m=1$.} 
\label{fig:fig6}
\end{figure}

For a resistive current-driven mode (a tearing mode) 
since the growth 
rate scales as $S^{-3/5}$ or $\eta^{3/5}$, where 
S is the Lundquist number,  
the Maxwell stress term is small in the outer region ($\eta=0$)
 and the main contribution
arises from the inner layer solution (resistive region).
Equation~\ref{eq:maxwell} only presents the outer solution and is singular 
around the reconnection layer (F=0). To obtain 
the radial structure of stresses for a tearing mode with a 
reconnecting surface $\textbf k \cdot \textbf B= F =0$, the inner layer equations
need to be solved (Ebrahimi, Mirnov, \& Prager 2008). 
We therefore perform nonlinear single tearing mode computations with two set of equilibria ($\lambda=10$ and $\lambda=17$).
The radial structure of the quasilinear Maxwell
and Reynolds stress terms during the linear phase for two cases 
are shown in Fig.~\ref{fig:fig6}. The mode structure for the case
with $\lambda_{max}=10$ is a single mode tearing structure. 
As discussed, the main contribution 
of MHD stresses for the tearing mode 
arises in the inner region, which is also confirmed by the 
computations [Fig.~\ref{fig:fig6}(a)]. As is seen in
Figure~\ref{fig:fig6}(a), the MHD stresses are 
localized around the reconnecting surface ($r \approx 0.6$), and are small in the outer 
ideal region.  Because the computation is in the visco-resistive regime, 
the Maxwell stress is much larger than 
the Reynolds stress. The localization of the stresses leads to 
a localized flattening of the azimuthal flow within the resistive layer. It is also seen in Fig.~\ref{fig:fig6}(a) that the 
azimuthal Lorentz force ($<\widetilde J \times \widetilde B>_{\phi}$) is bi-directional and transports momentum outward.  
As the free energy for the instability increases, for the larger
current ($\lambda_{max}=17$) case, double tearing 
mode become unstable which has two reconnecting surfaces 
at two radii ($r \approx 0.5$, $r \approx 0.8$).
The radial structure of the Maxwell stress shown in Fig.~\ref{fig:fig6}(b)
 is broader than the single tearing 
mode case. 

It should be mentioned that a more generalized set of ideal compressible equations 
including equilibrium flows 
in a cylindrical current-carrying plasma (Bondeson, Iacono \& Bhattacharjee 1987) and 
the gravitational force (Keppens, Casse \& Goedbloed 2002; Blokland, et al. 2005) 
have been obtained.
 They used Frieman \& Rotenberg (1960) 
formalism to obtain first order differential equations for displacement 
$\xi_r$ and $\Pi$ (total kinetic and magnetic pressure perturbations). 
Using these generalized equations (Eqs. 14-20 in Blokland et al. 2005), 
after some algebra 
we have also obtained the parallel magnetic field perturbation ($\widetilde{\textbf{B}} \cdot \textbf{B}/B$)  
which reduces to Eq.~\ref{eq:bbout} in the incompressible limit. 
We therefore, find that in the incompressible regime 
by ignoring the pressure perturbation (and with uniform equilibrium pressure),
but including the gravitational force perturbation, 
the resulting Maxwell stress term remains the same as Eq.~\ref{eq:maxwell}.

We also note that in a current-free plasma with a uniform axial magnetic field and 
a differential rotation, Eqs. \ref{eq:vrout} and \ref{eq:vrout2}  
reduce to
a differential equation for a global axisymmetric m=0 MRI (Velikhov 1959 
and Chandrasekhar 1960),
\begin{equation}
\frac{\partial ^2 \widetilde{V}_{r}}{\partial r^2}+ \frac{1}{r}\frac{\partial \widetilde{V}_{r}}{\partial r} - [\frac{1}{r^2} + k^2 + \frac{2r \Omega^{\prime}\Omega k^2}{(\omega_A^2-\gamma^2)} - \frac{4\Omega^2 \gamma^2 k^2}{(\omega_A^2-\gamma^2)^2}]\widetilde{V}_{r}=0
\label{eq:ode1} 
\end{equation} 
where, $\Omega= V_{\phi}(r)/r$, $\textbf{B}=B_0 \textbf{e}_z$, and 
$\omega_A^2 = k^2 B_{0}^2/\rho$.
It can be shown that the stresses from an axisymmetric MRI can
be global and depend on flow shear (Ebrahimi, Prager \& Schanck 2009)\nocite{ebrahimi2009}.

\section{{\Large\textbf{Multiple mode computations }}}
\label{sec:nonlin}
Here, we investigate full nonlinear dynamics of momentum transport from
current-driven instabilities in a disk geometry using multiple tearing 
mode computations.
In the nonlinear computations,
 with multiple modes (in both azimuthal 
and vertical directions), the 
additional effect of nonlinear mode coupling is revealed. 
Moreover, the transfer 
of energy from fluctuations to the mean field can occur during the nonlinear 
saturation. We perform full nonlinear computations when the 
current-driven modes with different azimuthal and axial 
mode numbers are included 
in the computations. Current-driven 
modes can nonlinearly interact and cause nonlinear
growth. Three sets of nonlinear computations are performed in a disk geometry.
The first two multiple mode computations are weakly nonlinear driven and discussed 
in section \ref{sec:nonlin1}.
 A multiple mode computation, which is strongly nonlinear driven, 
is presented in section \ref{sec:nonlin2}. In the weakly nonlinear driven case,
 many current-driven modes 
are linearly unstable while the effect of nonlinear coupling to stable modes is weak.
 In the strongly nonlinear case however, many stable modes are nonlinearly driven by
linearly unstable modes.

\subsection{{\Large\textbf{Weakly nonlinear driven}}}
\label{sec:nonlin1}
Here, two sets of computations are performed in a disk geometry with similar 
force-free equilibrium shown in Fig.~\ref{fig:fig2} with (1) $\lambda_{max} =9$
and (2) $\lambda_{max}=14.2$.  
Both computations start with a Keplerian flow 
profile with an on axis amplitude of $V_{\phi}/V_A=0.8$ 
(with $P_m=1$, $S=10^4$, $\beta_0=10$, 
 and radial, azimuthal and axial 
resolutions  $n_r$=220, $0<m<21$ and $-43<n<43$, respectively). 

\begin{figure} 
\includegraphics[]{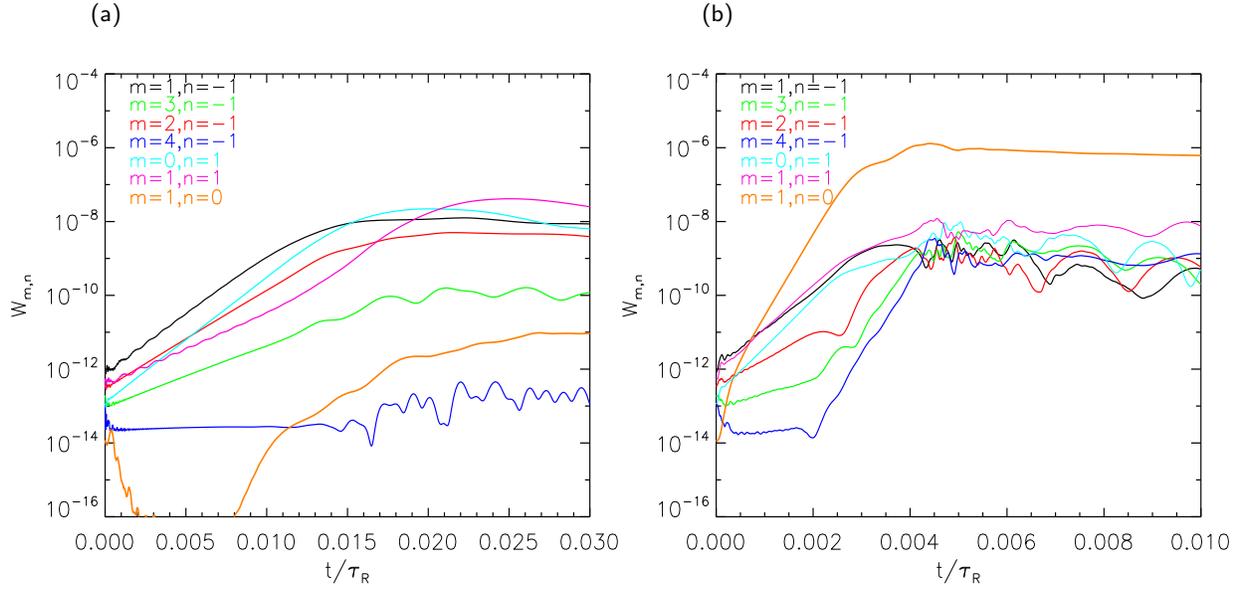}
\caption{ Magnetic energy $W_{m,n}=1/2 \int{\widetilde B_{r(m,n})^2 dr^3}$ vs time for
different tearing modes (m,n) for (a) $\lambda_{max}=9$ and (b) $\lambda_{max}=14.2$ }
\label{fig:fig7}
\end{figure}

The computations start with a current-carrying equilibrium, 
and the free energy from the parallel current causes the 
current-driven instabilities to grow. 
The radial magnetic energy for different current-driven tearing modes are shown in 
Fig.~\ref{fig:fig7}.
Tearing modes which are linearly unstable (with the growth rates given 
in Table 1 for equilibria with $\lambda_{max}=9$ and $\lambda_{max}=14.2$), 
are also driven linearly in the nonlinear computations. Moreover,
 other modes with higher azimuthal mode numbers are driven nonlinearly. 
As seen in Fig.~\ref{fig:fig7}(a), 
modes with tearing parity  m=0,1,2 (both n=1 and n=-1) linearly start to grow and saturate 
around t=0.015 $\tau_R$. For tearing mode m=1,n=1, there is a second nonlinear 
growth before saturation. Modes with higher azimuthal mode numbers also shown, 
and is seen that m=3 mode linearly grow with small growth rate.
 This mode is linearly stable in the absence of 
flow and here with the azimuthal flow becomes linearly unstable.
 For the case with equilibrium $\lambda_{max}=9$, the current gradient 
(free energy) is not large enough to make the ideal current-driven modes (including
non-resonant) unstable. Thus for this equilibrium, only resonant 
resistive tearing modes are unstable.   

When the free energy, i.e current gradient, increases both resistive 
and ideal current-driven modes becomes unstable. Figure \ref{fig:fig7} (b) 
shows the radial magnetic energies for several current-driven modes with the 
equilibrium $\lambda_{max}=14.2$. As seen, tearing modes m=0,1,2,and 3
linearly grow as expected from linear stability analysis (Table 1).
These tearing modes also have a nonlinear growth around t= 0.002$\tau_R$. 
 In addition to linearly unstable modes, m=4 tearing mode which is linearly 
stable, nonlinearly starts to grow around t=0.002$\tau_R$. 
For this equilibrium, as shown in Fig~\ref{fig:fig7}(b) 
ideal current-driven mode m=1 ,n=0 mode is linearly unstable
and saturate about two order of magnitude higher amplitude.   

\begin{figure}
\includegraphics[]{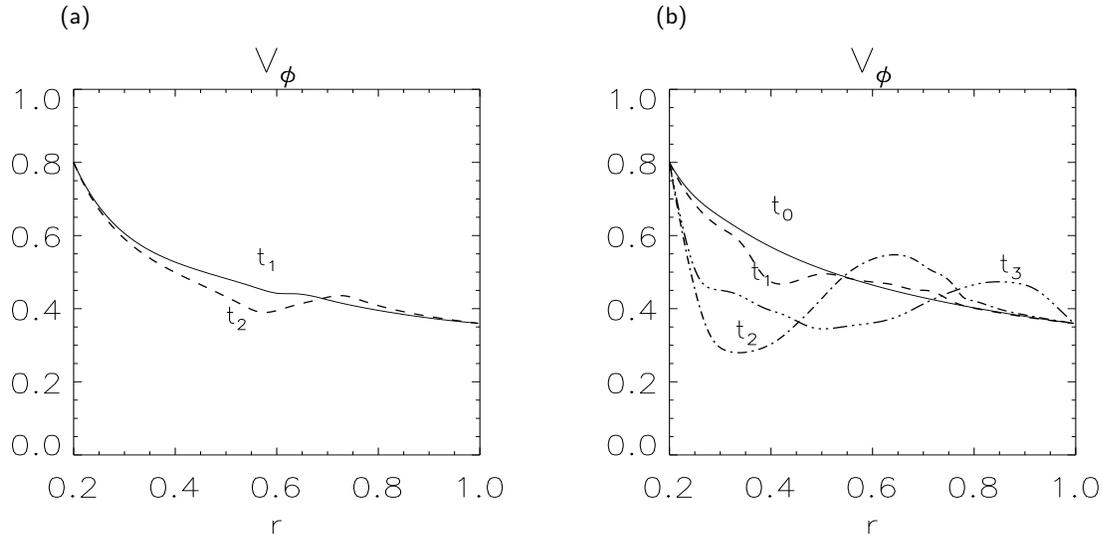}
\caption{.The evolution of flow profile during nonlinear computation for  
(a)  $\lambda_{max}=9$ at two times  $\mathrm t_1$=0.012,  $\mathrm t_2$=0.023 and (b) 
$\lambda_{max}=14.2$ at three times  $\mathrm t_1$=0.0028,  $\mathrm t_2$=0.0038, and  $\mathrm t_3$=0.0056.} 
\label{fig:fig8}
\end{figure}

\begin{figure} 
\includegraphics[]{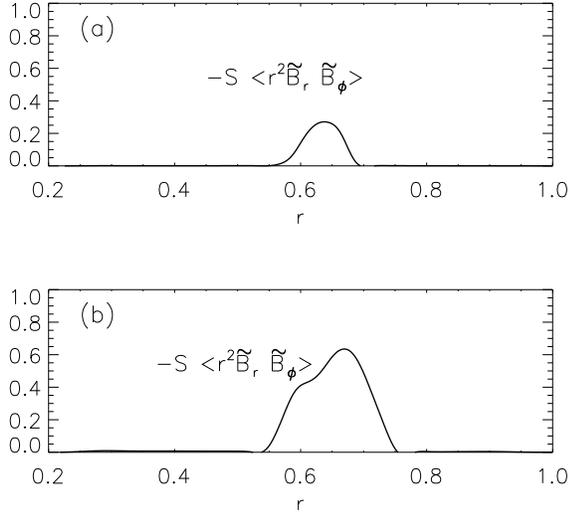}
\caption{ Radial profiles of total (surface-averaged including all the modes)
 Maxwell and Reynolds stress terms at two times (a) $\mathrm t_1$=0.012, (b) 
$\mathrm t_2$=0.023 ($\lambda_{max}$=9)}
\label{fig:fig9}
\end{figure}

As current-driven modes grow and nonlinearly saturate through modifying the 
source of instability, i.e the current gradient, they also modify 
the mean azimuthal 
flow profile. Figure \ref{fig:fig8} shows the modification of 
the azimuthal flow profile during the nonlinear evolution for both cases 
  $\lambda_{max}=9$ and $\lambda_{max}=14.2$. For the case with 
equilibrium  $\lambda_{max}=9$, the modification of flow profile is mainly due to the 
resistive tearing modes. The azimuthal flow profile at two times 
$\mathrm t_1=0.012$, before saturation, and $\mathrm t_2= 0.023$, after the saturation of
tearing modes, 
 are shown in Fig.~\ref{fig:fig8}(a) with $\lambda_{max}=9$. At t=$\mathrm t_1$, the flow is modified 
around r=0.6 due to m=1 tearing mode. As discussed before, 
 the azimuthal flow is modified through the Maxwell and Reynolds stresses.
The structure of the fluid stresses are shown in Figure \ref{fig:fig9}. 
The Maxwell stress at t=$\mathrm t_1$ is localized around r=0.6 and cause the modification
of the flow and transport of momentum outward.    
After the nonlinear saturation, the Maxwell stress from all the tearing modes
  becomes broader (Fig.~\ref{fig:fig9}(b)), thus cause a broader 
flow modification as shown at t=$\mathrm t_2$ Fig.~\ref{fig:fig8}(a).   
The structure of total nonlinear Maxwell stress (from multiple tearing 
computation) is broader than the structure for a single tearing mode
(Fig.~\ref{fig:fig9}(a)). The structure is more concentrated in the 
plasma core (around r=0.5 to r=0.8), where the current gradient is large. 

\begin{figure}
\includegraphics[]{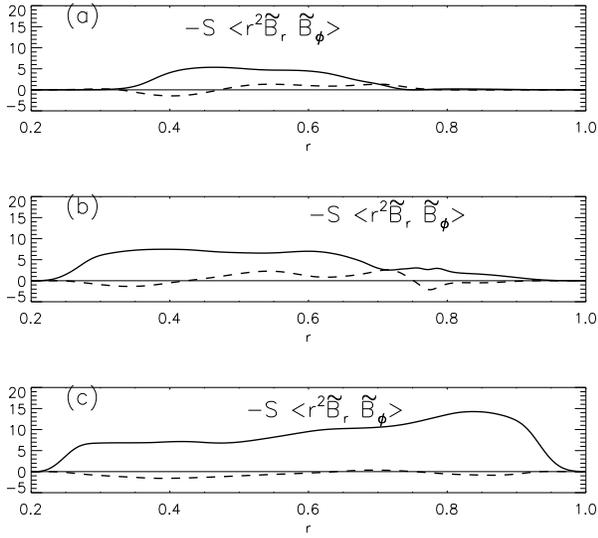}
\caption{Radial profiles of total Maxwell (solid lines) and Reynolds stress (dashed lines) terms at three times (a) $\mathrm t_1$=0.0028, (b) $\mathrm t_2$=0.0038, and (c) $\mathrm t_3$=0.0056, $\lambda_{max}$=14.2.}
\label{fig:fig10}
\end{figure}

The momentum transport from current-driven modes becomes stronger
when ideal modes are also present. The nonlinear evolution of 
azimuthal flow profile for the case with larger current gradient $\lambda_{max}=14.2$
is shown in Fig.~\ref{fig:fig8}(b). The structure of Maxwell and Reynolds stress
terms are also shown in Fig.~\ref{fig:fig10}. As seen, 
the Maxwell stress
is concentrated in the core region at t=$\mathrm t_1$ before the nonlinear saturation
(Fig~\ref{fig:fig10}(a)) and the flow modification also occurs around the 
region where the stresses are peaked from r=0.4 to r=0.7. Around the nonlinear
saturation at t=$\mathrm t_2$ and after the saturation at t=$\mathrm t_3$, it is shown that 
fluid stresses become more global (Fig.~\ref{fig:fig10}(b),(c))
 and cause a global modification of the 
azimuthal flow profile and momentum transport outward (Fig.~\ref{fig:fig8}(b)).   
The global modification is due to both resistive tearing and ideal current-driven modes.

It is also worth estimating the effectiveness of the 
current-driven instabilities for angular momentum transport using the standard 
Shakura-Sunyaev $\alpha$ model. 
In this model the effective viscosity is parametrized by the quantity 
$\alpha_{SS}$ which relates to the Reynolds and Maxwell stress terms by 
$<\rho \widetilde V_r \widetilde V_{\phi}> - <\widetilde B_r \widetilde B_{\phi}> = \alpha_{SS} <P>$, where P is the average pressure. 
The normalized alpha used for our simulations is $\alpha_{SS}/\beta_0/2$.
For the two weakly nonlinear driven cases 
presented here, $\lambda_{\mathrm {max}}$ =9 and $\lambda_{\mathrm {max}}$=14.2,  the 
time-averaged Shakura-Sunyaev $\alpha$, $<\alpha_{SS}>$, during nonlinear saturation are
$<\alpha_{SS}> \approx 5 \cdot 10^{-6}$ and  
$<\alpha_{SS}> \approx 4 \cdot  10^{-4}$, respectively. We have also calculated, the 
time-averaged Shakura-Sunyaev $\alpha$ for a weakly nonlinear case 
with $\lambda_{\mathrm {max}}$=14.2 and $\beta_0=1$,
 which is $<\alpha_{SS}> \approx 2.5 \cdot 10^{-3}$.

\subsection{{\Large\textbf{Strongly nonlinear driven}}}
\label{sec:nonlin2}

\begin{figure} 
\includegraphics[]{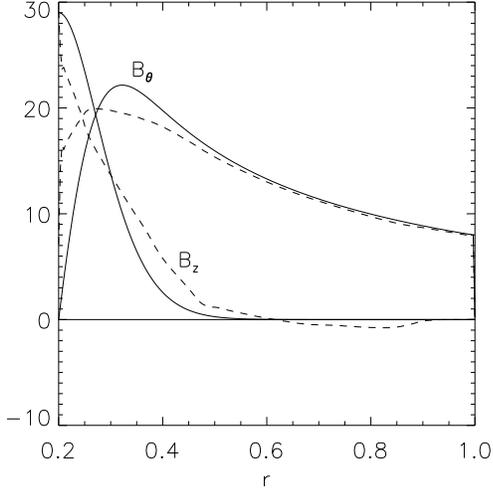}
\caption{Equilibrium azimuthal and vertical magnetic field profiles for strongly nonlinear driven case
at t=0 (solid lines) and at t=0.0062 during nonlinear saturation (dashed lines).}
\label{fig:fig11}
\end{figure}

\begin{figure} 
\includegraphics[]{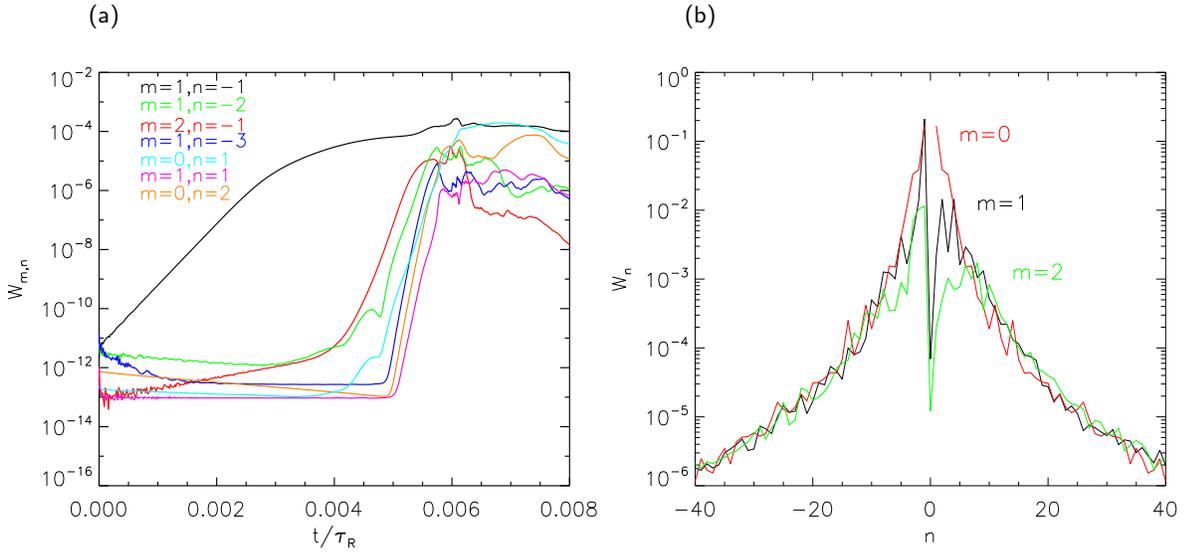}
\caption{(a) Magnetic energy $W_{m,n}=1/2 \int{\widetilde B_{r(m,n})^2 dr^3}$ vs time for
different tearing modes (m,n), (b) magnetic energy spectrum; for the strongly nonlinear driven case.}
\label{fig:fig12}
\end{figure}

We further investigate the effect of nonlinear mode coupling in a different equilibrium setting.
The computations are performed in a force-free equilibrium for which the 
components of mean magnetic fields are shown in Fig.~\ref{fig:fig11} (solid lines). 
   Full nonlinear computation starts with a Keplerian flow 
profile with an on axis amplitude of $V_{\phi}/V_A=0.8$ 
(with $P_m=5$, $S=10^4$, $\beta=10$, 
 and radial, azimuthal and axial 
resolutions  $n_r$=250, $0<m<11$ and $-43<n<43$, respectively). 
The free energy, parallel current, is mainly concentrated in the inner half  plasma 
region and causes a current-driven instability with mode number m=1, n=-1 to 
become linearly unstable.
The radial magnetic energy for different current-driven tearing modes is shown in 
Fig.~\ref{fig:fig12}. Almost all the modes are stable up to around 
$\mathrm t/\tau_R=0.004$ and plasma is in a quasi-single mode state
 (with m=1, n=-1).
Two other tearing modes, m=1, n=-2 and m=2, n=-1 also grow linearly with small growth rates.
As seen in Fig.~\ref{fig:fig12}(a), around $\mathrm t/\tau_R=0.004$, 
due to nonlinear mode coupling axisymmetric m=0 mode becomes nonlinearly 
unstable and saturates at a  large amplitude comparable to the amplitude of 
the initial m=1 linearly-driven
 mode.  Other non-axisymmetric modes are also driven nonlinearly and a 
turbulent state is formed. The magnetic spectrum for m=0--2 
during the nonlinear state 
is shown in Fig.~\ref{fig:fig12}(b), as can be seen, the spectrum 
shows a broad range of 
magnetic fluctuations. The transition from a quasi-single mode
to a multiple-mode state occurs due to strong mode-mode coupling and transfer
of energy from m=1 modes to m=0 modes.
In the weakly nonlinear driven case (section~\ref{sec:nonlin1}), the equilibrium 
was linearly unstable for most of the current-driven modes, and the nonlinearly driven modes  
(m=3 and 
m=4 modes for the cases with $\lambda_{max}=9.$ and $\lambda_{max}=14.2$, respectively) would not grow to large amplitudes. Here most of the tearing modes are
 driven nonlinearly and saturate at large amplitudes.

\begin{figure} 
\includegraphics[]{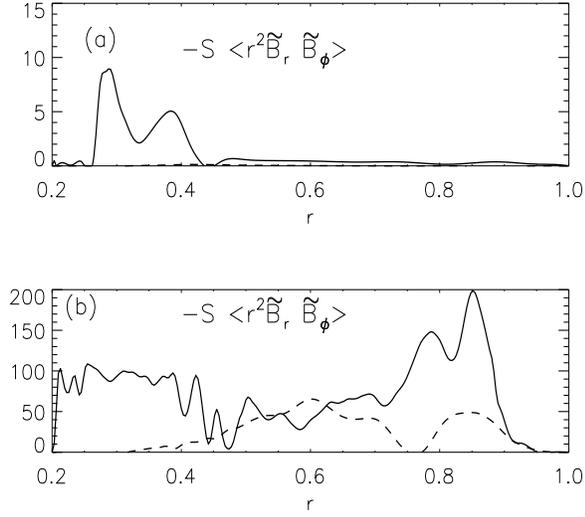}
\caption{Radial profiles of total Maxwell (solid lines) and Reynolds stress (dashed lines) terms (a) at $\mathrm t_1=0.0047$ 
during the single mode state (b) at $\mathrm t_2=0.0062$ during the nonlinear multiple mode state.}
\label{fig:fig13}
\end{figure}

Tearing fluctuations during the nonlinear state ($\mathrm t/\tau_R>0.005$) can affect the 
mean profiles through the fluctuation-induced convolutions terms. 
Mean magnetic fields during the nonlinear saturation, which are 
affected by the magnetic fluctuations, are shown in Fig.~\ref{fig:fig11}
 (the dashed lines). As it is seen, the 
vertical magnetic field changes sign and toroidal flux is redistributed by the 
fluctuation-induced term, $<\widetilde V \times \widetilde B>$, 
the so called dynamo term. The tearing fluctuations also affect
the flow profile and cause momentum transport. The radial structures of 
Maxwell and Reynolds stresses during the two states, quasi-single mode and multiple
modes, are shown in Fig~\ref{fig:fig13}. 
During the quasi-single mode state ($\mathrm t/\tau_R<0.004$),
the Maxwell stress transports momentum outward, but it 
is localized in the inner region 
around the reconnection layer. However, during the 
multiple-mode state ($\mathrm t/\tau_R>0.005$)        
the structure of total stresses is very global and all the 
tearing modes contribute 
to the momentum transport. It can also be seen that the Maxwell 
stress is much stronger (about an order of magnitude)
for the strongly nonlinear driven case compared to the weakly 
nonlinear driven case (Fig.~\ref{fig:fig10}).

\begin{figure} 
\includegraphics[]{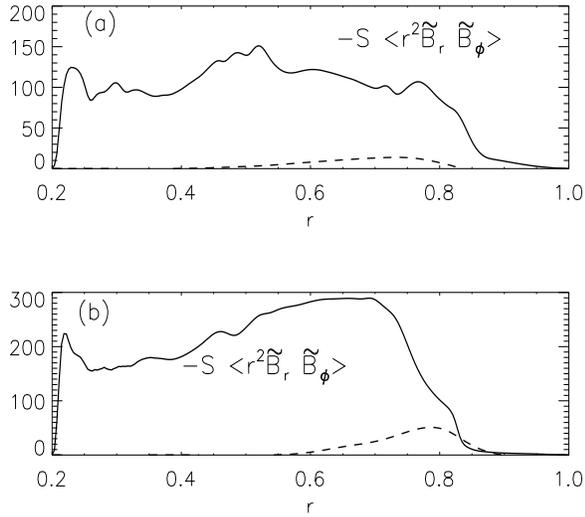}
\caption{Radial profiles of total Maxwell and Reynolds stress terms after nonlinear 
saturation for (a) $\beta_0$ = 10 (b) $\beta_0$ =1; Pm =20, S=$10^4$. }
\label{fig:fig14}
\end{figure}

We have also performed strongly nonlinear driven cases with the same force-free 
equilibrium shown in Fig.~\ref{fig:fig11} but 
 with Pm=20, $\beta=10$, and Pm=20, $\beta=1$. The structure of stresses after 
nonlinear saturation (after the growth and saturation  of nonlinearly driven mode, m=0) 
are shown in Fig.~\ref{fig:fig14}. As seen,
 Maxwell stress is broad due to nonlinear mode coupling. We have calculated 
the
time-averaged Shakura-Sunyaev $\alpha$ during the nonlinear saturation, which is 
$<\alpha_{SS}> \approx 4.3 \cdot 10^{-3}$ and $<\alpha_{SS}> \approx 3.3 \cdot 10^{-2}$ for the two cases 
with  $\beta=10$ and $\beta=1$, respectively.

\section{{\Large\textbf{Summary}}} 
\label{sec:summ}
We have shown that tearing instabilities can transport 
angular momentum effectively in thick disks.  A single 
tearing mode transports momentum locally, in the 
vicinity of the radial reconnection location.  
However, with multiple modes simultaneously unstable 
and nonlinearly coupled to each other, the resultant 
Maxwell stress and momentum transport can span the 
entire plasma cross-section. 
It is also shown that an ideal current-driven instability without a reconnecting 
surface has 
a global Maxwell stress and therefore can produce more global momentum transport.
However, the global Maxwell stress from  many tearing modes causes 
much stronger momentum transport compared to an ideal 
current-driven instability with a global stress.
These conclusions are 
drawn from a hierarchy of calculations, including 
linear stability, quasilinear analytic calculations, 
and computation of the full nonlinear evolution 
of multiple modes.

These instabilities, although present in
rotating plasmas, are driven magnetically by the 
plasma current. We have selected for study disk equilibria 
for which the magnetic field is sufficiently 
strong that the plasma is stable to the flow-driven 
magnetorotational instability.  Determination of 
the relevance of tearing instabilities to 
astrophysical disks requires extension of these 
studies to a wider range of conditions, such as to
thin disks and different magnetic field strength.
As the disk becomes thinner, the relevant tearing modes (with smaller wavelengths)
may become more stable. The effectiveness of tearing modes in 
transporting momentum in disks 
also depends on the ratio of magnetic energy to the flow energy. 
Although in the linear regime, ideal current-driven instabilities
 are independent of resistivity, 
nonlinear momentum transport from both tearing and ideal current-driven 
instabilities in the low magnetic diffusivity regime could be an interesting 
topic for a future work. \\

\acknowledgments
The authors wish to thank useful discussions with Dalton Schnack, Ellen Zweibel and 
Amitava Bhattacharjee.\\

\begin{thebibliography}{14}
\expandafter\ifx\csname natexlab\endcsname\relax\def\natexlab#1{#1}\fi

\bibitem[{Balbus \& Hawley(1991)}]{balbus91}
Balbus, S.~A., \& Hawley, J.~F. 1991, \apj, 376, 214

\bibitem[{Balbus \& Hawley(1998)}]{balbus98_rmp}
---. 1998, Rev. Mod. Phys., 70, 1

\bibitem[{Balbus \& Terquem(2001)}]{balbus2fl}
Balbus, S.~A., \& Terquem, C. 2001, Astrophys. J., 552, 235

\bibitem[{Biskamp(1993)}]{biskamp}
Biskamp, D., 1993, Nonlinear magnetohydrodynamics, vol. 1 of Cambridge Monographs on Plasma Physics, Cambridge (Mass.): CUP

\bibitem[{Blokland et al. (2005)}]{blokland}
Blokland, J. W. S., van der Swaluw, E., Keppens, R., \& Coedbloed, J. P. 2005, A\&A, 444, 337

\bibitem[{Bondeson {et~al.}(1987)Bondeson, Iacono, \& Bhattacharjee}]{bondeson}
Bondeson, A., Iacono, R., \& Bhattacharjee, A. 1987, Phys. Fluids, 30, 2167

\bibitem[Brandenburg et al. (1995)]{branden05}
Brandenburg, A., Nurdlund, A., Stein, R. F., \& Torkelsson, U. 1995, \apj, 446, 741

\bibitem[{Chandrasekkar(1961)}]{chandrasekkhar61}
Chandrasekkar, S. 1961, Hydrodynamic and hydromagnetic stability (Dover)

\bibitem[{Donati {et~al.}(2007)Donati, Jardine, Gregory, Petit, Bouvier, \&
  et~al}]{donati07}
Donati, J.-F., Jardine, M. M.~., Gregory, S.~G., Petit, P., Bouvier, J., \&
  et~al. 2007, MNRAS, 380, 1297

\bibitem[{Donati {et~al.}(2005)Donati, Paletou, Bouvier, \&
  Ferreira}]{donati05}
Donati, J.-F., Paletou, F., Bouvier, J., \& Ferreira, J. 2005, Nature, 438, 466

\bibitem[{Ebrahimi {et~al.}(2009)Ebrahimi, Mirnov, \& Prager}]{ebrahimi2008}
Ebrahimi, F., Mirnov, V. V.,  \& Prager, S.~C. 2008, Phys of Plasmas, 15, 055701

\bibitem[{Ebrahimi {et~al.}(2009)Ebrahimi, Prager, \& Schnack}]{ebrahimi2009}
Ebrahimi, F., Prager, S.~C., \& Schnack, D.~D. 2009, \apj, 698, 233

\bibitem[{Fleming \& Stone (2003)}]{fleming} 
Fleming, T., \& Stone, J. 2003, \apj,, 585, 908

\bibitem[{Frieman \& Rotenberg (1960)}]{frieman} 
Frieman, E., \& Rotenberg, M. 1960, Rev. Mod. Phys., 32, 898

\bibitem[{Furth {et~al.}(1963)Furth, Killeen, \& Rosenbluth}]{fkr}
Furth, H.~P., Killeen, J., \& Rosenbluth, M.~N. 1963, Phys. Fluids, 6, 459

\bibitem[{Gammie(1996)}]{gammie96} 
Gammie, C.~F. 1996, \apj, 457, 355

\bibitem[{Goodman(2003)}]{goodman} 
Goodman, J. 2003, MNRAS, 339, 937

\bibitem[Hawley et al. (2001)]{hawley01}
Hawley, J. F., Gammie, C. F., \& Balbus, S. A. 2001, \apj, 554, 534

\bibitem[{Ji {et~al.}(2006)Ji, Burin, Schartman, \& Goodman}]{ji06}
Ji, H., Burin, M., Schartman, E., \& Goodman, J. 2006, Nature, 444, 343

\bibitem[Keppens et al. (2002)]{keppens}
Keppens, R., Casse, F., \& Goedbloed, J. P. 2002, \apj, 569, L121

\bibitem[{Levy (1978)}]{levy78} 
Levy, E. H. 1978, Nature, 276, 481

\bibitem[{Levy \& Araki(1989)}]{levya} 
Levy, E. H., \& Araki, S. 1989, Icarus, 81, 74

\bibitem[{long et al. (2008)}]{long} 
Long, M., Romanova, M. M., \& Lovelace, R. V. E. 2008, Mon. Not. R. Astron. Soc., 386, 1274 

\bibitem[{Miller \& Stone(2000)}]{miller00}
Miller, K., \& Stone, J. 2000, Astrophys.J., 534, 398

\bibitem[{Newcomb(1960)}]{newcomb}
Newcomb, W.~A. 1960, Ann. Phys. (N.Y.), 10, 232

\bibitem[{Pringle(1981)}]{Pringle81}
Pringle, J.~E. 1981, Ann. Rev. Astron. Astrophys., 19, 137

\bibitem[{Reyes-Ruiz et al. (1999)}]{reyes}
Reyes-Ruiz, M., \& Stepinski, T. F. 1999, A\&A, 342, 892

\bibitem[{Rudiger et al. (1995)}]{Rudiger}
Rudiger, G., Elstner, D., \& Stepinski, T. F. 1995, A\&A, 298, 934

\bibitem[{Shakura \& Sunyaev(1973)}]{shakura73}
Shakura, N.~I., \& Sunyaev, R.~A. 1973, Astron. Astrophys, 24, 337

\bibitem[{Stepinski \& Levy(1990)}]{stepinski}
Stepinski, T. F., \& Levy, E.~H. 1990, \apj, 350, 819

\bibitem[{Torkelsson \& Brandenburg(1994)}]{torkelsson}
Torkelsson, U., \& Brandenburg,~A. 1994, Astron. Astrophys, 283, 677

\bibitem[{Uzdensky(2008)}]{Uzdensky}
Uzdensky, D. A., \& Goodman, J. 2008, \apj, 682, 608

\bibitem[{Velikhov(1959)}]{velikhov59}
Velikhov, E.~P. 1959, Sov. Physics JETP, 36, 995

\bibitem[{Wardle(1999)}]{wardle}
Wardle, M. 1999, Mon. Not. R. Astron. Soc., 307, 849

\bibitem[{Wardle(2007)}]{wardle}
Wardle, M. 2007, Astrophys Space Sci, 311, 35

\bibitem[{Zhu {et~al.}(2010)Zhu, Hartmann, \& Gammie}]{gammie10}
Zhu, Z., Hartmann, L., \& Gammie, C.~F. 2010, \apj,, 713, 1143

\end{thebibliography}
\end{document}